\documentclass{aastex}

\begin{document}

\title{The Large Magellanic Cloud and the Distance Scale\\ }
\shorttitle{LMC Distance}
\shortauthors{Walker, Alistair R.}

\author{Alistair R. Walker\altaffilmark{1}}
\affil{Cerro Tololo Inter-American Observatory, National Optical Astronomy Observatory, Casilla 603,
La Serena, Chile.}
\email{awalker@ctio.noao.edu}

\begin{abstract}

The Magellanic Clouds, especially the Large Magellanic Cloud, are places where
multiple distance indicators can be compared with each other in a straight-forward manner
at considerable precision.   We here review the distances derived from Cepheids, Red 
Variables, RR Lyraes, Red Clump Stars and Eclipsing Binaries, and show that
the results from these distance indicators generally agree to within their errors,
and the distance modulus to the Large Magellanic Cloud appears to be defined to $\pm3\%$ with
a mean value  $(m-M)_{0} = 18.48$ mag, corresponding to 49.7 Kpc.   
The utility of the Magellanic Clouds in constructing 
and testing the distance scale will remain as we move into the era of Gaia.

\end{abstract}



\section{Introduction}

The Magellanic Clouds (MC) are the closest populous galaxies to our own, and as such contain 
significant numbers of most of the indicators used to make up the distance ladder 
that transfers geometrically measured distances for nearby stars to the far universe via multiple 
overlapping steps.   Both galaxies are now realized to be complex systems interacting 
with each other but not perhaps as-yet with the Galaxy \citep{nid10,ols11,bes11}.

The main structures of the Large Magellanic Cloud (LMC) lie reasonably close to the plane 
of the sky \citep{nik04}, but the Small Magellanic Cloud (SMC)
 has long been realized \citep{cal86} to form an extended structure 
almost in the line of sight.  Although
at times a Galaxy-LMC-SMC comparison is useful to test the
metallicity dependence of some particular distance indicator, in general the SMC is of rather less 
utility than the LMC for distance scale work.  The approximation that the LMC is sufficiently compact and distant that its 
contents are all at the same distance from us - and yet close enough that
crowding and faintness are usually not issues - is a highly valuable attribute, although when pushing for 
distance comparisons at the few percent level the effects of sample size and spatial distribution
will be important, and corrections for the 
geometry of the LMC, should be carefully considered.    For small samples the possible line 
of sight variation in distance to any particular star must be taken into account as it
may be significantly different from the mean distance, and for unique 
distance indicators such as SN 1987A the distance determined is of course that for the object itself and to 
go from that distance to the mean distance of the corresponding galaxy requires an extra step with 
its own associated error.  In some cases the latter may be difficult to determine \citep{sah10,ols11}.

Over the past decade the controversy over the ''long versus short'' distance scale  
has largely been resolved, with the corresponding values of the Hubble Constant 
$(\sim 50$ vs. $\sim 100$ 
respectively), settling to an intermediate value, with rather small errors usually
 quoted \citep{fre10,rie11}.
Although the range of LMC distances given by all available indicators in an oft-quoted
compendium by \citet{ben02} is large, some of the more notable outliers (e.g. RGB clump stars) 
are now better understood and more nuanced evaluations have generally been 
consistent for the distance indicators considered most reliable \citep{wal03,alv05,cle08}, 
to ten percent or better. 
  Does this mean that the distance scale is now rather a less important 
topic than when apparently careful studies obtained values for the Hubble Constant that were a 
factor-of-two different?  Today in the era of precision cosmology, as typified by the 
Wilkenson Microwave Anisotropy Probe (WMAP) results \citep{kom11},
the Hubble Constant is generally used as a prior in the interpretation of
the relations that describe the components that make up the energy-density of the Universe.
Knowing the Hubble Constant to a (very) few percent 
is thus essential \citep{fre10} and so we conclude that the study of distance indicators is 
still of significant scientific importance.   

The improvement of instrumentation over the past decade and a half, chiefly the availability
of the Hubble Space 
Telescope (HST) with its series of imaging cameras (WFPC2, ACS, WFC3), has allowed the 
Cepheid distance scale to be extended well beyond the Local Group. Additionally,
measurements with the Fine Guiding Sensors (FGS) have allowed
a direct calibration by parallaxes for 10 Galactic Cepheids \citep{ben07}, and with a 
geometrical distance to the maser galaxy ''NGC 4258'' \citep{mac06} 
available for comparison, the ''one step'' jump to the Supernovae Ia (SNIa)  that are used 
to probe to great distances would seem to reduce the importance of the MC
in distance 
scale work.  But the history of the subject shows systematic errors are invidious; 
the well known adage that ''The Hubble 
Constant at any given time has always been known to 10 percent, despite having changed over 
that period by a factor of 10'' should not be forgotten. For these reasons the importance of 
the MC, and especially the LMC remains intact,
providing a sanity check on the validity of the lower rungs of the distance scale ladder.   
The availability of large telescopes with relatively wide-field imagers (e.g.
 Subaru, Magellan, LBT) means that such comparisons are not restricted to the MC, with the
 more luminous distance indicators able to be compared for instance in M31 and M33.

 The discussion here is 
not meant to be inclusive of all distance indicators that have ever been suggested, and 
specifically excludes those that at the present time have insecure calibration or large errors.
The meeting associated with this review 
had as its focus {\it The Gaia Perspective} and thus the distance indicators chosen here:  Cepheids,
Red Variables, RR Lyraes, Red Clump Stars, Eclipsing binaries - all have their zero-point calibrations
depending on fundamental geometric methods. These calibrations will be greatly improved by Gaia, 
and some other distance indicators can be expected to join this list.   This 
selection does of course introduce some bias of its own, which the reader is well 
advised to bear in mind \citep{sha08}.  Unless stated otherwise the distances tabulated are
those determined by the authors of the papers quoted.  The recent comprehensive 
review of pulsating variable stars in the MC by \citet{cle08} 
 is highly recommended for a discussion complementary to that here; other important recent discussions 
are those of \citet{tam08,fre10} both of which contain copious references to earlier work.

\section{Cepheids}

The zero-point of the Cepheid period-luminosity (PL) relation (Leavitt Law) is now best determined
from the parallax measurements for 10 Cepheids determined directly with HST FGS 
\citep{ben02,ben07}.  The results can be combined with data for Cepheids in clusters
and associations \citep{few87,tur02} or with pulsation parallaxes (Baade-Wesselink method) 
e.g. \citet{fou07} to form accurate PL relations covering a wide period range \citep{tur10}.
The average error in the
individual HST Cepheid parallaxes is 8\%, it is noteworthy that these errors are typically a factor 10 smaller 
than those measured for 220 Galactic Cepheids by Hipparcos, as discussed by \citet{fea97}.  \citet{ben07}
compare their fitted galactic PL relations to those derived
from several studies and find that the slopes of the $K$ and Wesenheit $W(V,I)$ band passes are identical in
the Galaxy and the LMC to within the errors, and derive a $W(V,I)$ distance modulus $(m-M)_{0}$ of 
$18.50 \pm 0.03$, with no correction for any metallicity effect on the Cepheid absolute magnitudes. An 
analysis combining the two data sets (Hipparcos, HST) by \citet{van07} produces very similar results:
LMC moduli of $18.52 \pm 0.03$ ($W_{VI}$), $18.47 \pm 0.03$ ($K$ mag PL relation), and $18.45 \pm 0.04$ ($J,K$ mag 
period-luminosity-color relation), again with no correction for possible metallicity effects.  The
effectiveness in using directly measured Cepheid parallaxes is obvious, and the improvement that
will come with Gaia is enormous, with almost half the 20.000 estimated Cepheids in our Galaxy being
accessible \citep{win11}.

The concerns when transferring a Cepheid PL relation from the Galaxy to the LMC and beyond mostly are regarding the effect
of metal abundance on the slopes \citep{gie05,fou07,mad09} and zero-points \citep{bon08,kud08,sco09,sto11,fre11} of the PL relation.  Careful
choices of targets and band pass(es) can reduce degeneracies from reddening and period distribution differences,
and comparison of the Cepheid distance scale with other methods for galaxies with a wide range of environmental
conditions is instructive \citep{mou08,mou109,mou209,his11}. Treating this controversial topic \citep{sak04,rom09} 
in detail is beyond the scope of this review.   

 \citet{sto11, sto211} apply a newly revised calibration of
the infrared surface brightness technique to 41 MC stars (36 LMC, 5 SMC) and derive a distance modulus for the LMC 
of $18.46 \pm 0.04$ (statistical) where the zero-point is anchored to the Cepheids with HST parallaxes and thus carries 
its own uncertainty of another $\sim 5\%$, and where the projection ($p$) factor that converts observed radial velocities 
to pulsation velocities is constrained by being forced to show no variation in LMC distance as a function of pulsation period.
In particular, the $K$-band PL relation is found to have very little dependence on metallicity, in addition to the 
well-known advantages of lack of reddening sensitivity and less intrinsic scatter than PL relations at shorter wavelengths. 
This result is consistent with that found by \citet{fre09}.  \cite{fre10} provide two recent analyses of LMC Cepheid data, 
each referenced to the \citet{ben07} zero-point.  The first is a multi-wavelength solution based on the apparent $BVIJHK$ Cepheid 
distance moduli, fitting to the \citet{car89} extinction curve, to find that the data are well-fit by an LMC distance modulus 
of $18.40 \pm 0.01$ mag and $E(B-V) = 0.10$ mag.  The second method involves fitting the LMC Cepheids to the Galactic Cepheids with
parallaxes in the reddening-free Wesenheit $W(V,I)$ PL relation.  With no correction for metallicity the LMC modulus so derived is
$18.44 \pm 0.03$ mag, with a correction corresponding to 0.25 mag/dex \citep{sak04} the LMC modulus would decrease to $18.39$.
 
We conclude that almost all the recent work, when zero-pointed from the \citet{ben07} Galactic calibration, give LMC distance moduli in
the range $18.40-18.50$, with mean about $18.45$ mag.   Effects arising from the difference in metallicity between the the LMC 
and the Galaxy are the present major source of systematic error, 
although recent evidence hints at the effect of metallicity on Cepheid absolute magnitudes being very small, particularly in $K$ \citep{sto11}
and $W(V,I)$ \citep{maj11} band passes.  Additionally, both these band passes are relatively insensitive to the effects of reddening.

\section{Red Variables}

Miras are pulsating variable stars  on the Asymptotic Giant Branch (AGB) stage of their evolution, with 
progenitor masses in the range $ \sim 0.8 - 8 M_{\odot}$.   Those most useful for distance scale work have periods 
in the range $200-400$ days where tight PL relations are found, particularly in the $K$ band e.g. \citet{whi08}.  
Data from the rich MC samples in the newer surveys such as the OGLE III catalog \citep{sos09}, and 
cross identifying the stars in various IR catalogs show well-delineated PL relations.  Careful comparisons between 
measurements in the different band passes appears to allow separation of stars 
with different surface compositions (e.g. C stars), different evolutionary states (e.g. supergiants with longer
periods) and different pulsation modes.  The effects of circumstellar material on the band passes 
\citep{ita11} can be significant.  The large samples of stars also help to smooth out the ''noise'' of individual stars, which typically 
display 0.1 mag of secular variation in the $K$ band \citep{whi08}.

The analysis of the Hipparcos parallax data by \citet{whi00} was revised and extended by \citet{whi08} who also add
5 OH masers with parallaxes from the ESO VLTI, and 11 Miras in globular clusters where the distance zero-point is
based on Hipparcos parallaxes for subdwarfs \citep{car00}. The corresponding infrared photometry is almost 
exclusively from the SAAO. They divide their Hipparcos sample into carbon-rich and 
oxygen-rich samples, and then subdivide these samples in various ways, finding that carbon-rich stars give a similar 
zero-point but with larger uncertainty, and no significant effect with metallicity is found.   
Assuming the slope for the PL($K$) relation as
found in the LMC, their preferred solution for the O-rich galactic Miras is

\begin{displaymath}
M_{K} = -3.51\pm 0.20 (log P - 2.38) -7.25 \pm 0.07
\end{displaymath}

and thus, taking in quadrature their error of $\pm 0.06$ mag for the zero-point error of their fit to the LMC Miras, 
 a Mira distance modulus for the LMC of $18.49 \pm 0.09$ mag is obtained.

\citet{tab10} used a complementary approach in their analysis of $K$-band light curves for 247 southern, semi-regular 
red variables with Hipparcos parallaxes, allocating the stars to the several sequences in the $log P - M_{K}$ diagram
\citep{woo00,sos04} and comparing to stars in the MC.  They show the various P-L zero-points have 
a negligible metallicity dependence, and find an LMC modulus of $18.54 \pm 0.03$ (internal), plus the modulus difference
between the LMC and SMC is $0.41 \pm 0.02 $mag.

We conclude that the results from  the red variables give an LMC distance modulus of close to $18.51$ mag.
      
\section{RR Lyraes}
     
RR Lyraes are low mass ($\sim 0.6-0.8 M_{\odot}$) core-helium burning stars that lie within the pulsational instability strip 
in the horizontal branch region of the Hertzsprung-Russell (HR) diagram.  In the $V$ band the absolute magnitudes 
exhibit a dependence on metallicity that appears from both observations and theory to be linear over a wide range (at least 
$ -2.5 < [Fe/H] < -1.0$) and universal \citep{bon08,cle08}, with \citet{gra04} finding by combining their spectroscopy
of $\sim 100$ field RR Lyraes from the photometric study by \citet{cle03} that 

\begin{displaymath}
M_{V} = (0.214 \pm 0.047)([Fe/H] + 1.5) + constant.
\end{displaymath}

However, this simple relation may not be valid in some environments, for instance \citet{cap11} cautions that the the luminosity 
on the horizontal branch is very dependent on the helium abundance $Y$, with $\Delta M_{V} / \Delta Y  \sim -4.0$,
16 times the effect of metallicity $Z$.  Additionally, there is some evidence e.g. \citet{dic04} that at the higher 
metallicities the simple relationship with $M_{V}$ becomes non-linear, while evolutionary effects can intrude when comparing RR Lyraes
between globular clusters with differing distributions of horizontal branch stars.  In the near infrared (NIR) observational data are 
well fit by a period-luminosity-metallicity (e.g. $P-L_{K}-Z$) relation, with the advantages and disadvantages of this method 
very clearly summarized by \citet{cop11}.

The recently released OGLE-III catalog of 24906 RR Lyraes stars in the LMC is an enormous increase in the numbers known and apart from
the specific use as distance indicators the catalog is invaluable for studies of the LMC structure \citep{pej09}.  But even for earlier surveys
the number of stars are substantial, e.g. the OGLE-II catalog contains over 500 RR Lyrae stars for the SMC \citep{sos02}.  There are
also a few old clusters in the LMC, and one in the SMC (NGC 121) that contain RR Lyraes, see e.g. \citet{nem09}.

The zero-point calibration rests on the parallaxes for 
five RR Lyrae stars, a calibration via globular clusters whose distances are determined 
via absolute magnitudes for subdwarfs as measured by Hipparcos, and from various theoretical and empirical relations.
  \citet{cat08} 
form a weighted mean of the three parallax measurements available (HST, Hipparcos, Ground) for RR Lyrae itself to obtain a final value of
$\pi_{abs} = 3.78 \pm 0.19$ mas, corresponding to a distance modulus of $7.11 \pm 0.11$ mag.  By means of theoretical 
models relating absolute magnitudes for  RR Lyrae stars to their periods and the Str\"{o}mgren pseudo-color $c_{0}$ together 
with their calculated reddening and intensity-mean $V$ magnitude ($<V>$) they derive an absolute magnitude for 
RR Lyrae of $<M_{V}> = 0.600 \pm 0.126$.  Such careful analysis is essential when the calibration rests on a single star; the
situation has improved very recently with the publication of parallaxes for five RR Lyrae stars (including an improved parallax for RR Lyrae)
by \citet{ben11}, received as a preprint as this review was in preparation.

Distances to globular clusters can at present be most directly determined via fitting local
subdwarfs whose distances have been determined by Hipparcos to the unevolved main sequence of the GC in the CMD \citep{car00}. 
Such calibrations require the cluster reddening and metallicity as input, and for well-studied globular clusters the distances
so-obtained agree within the errors with other methods e.g. see \citet{cop11} for M5.  This results will soon be greatly
strengthened by improved HST FGS parallaxes for more subdwarfs \citep{cha11}. The NIR PL relations derived theoretically
by \citet{bon03} and \citet{cat04} also agree rather well with those using either of the above empirical methods.

There are a number of recent independent studies measuring RR Lyraes in the NIR in the LMC for both field and clusters \citep{bor04,
dal04, sze08}, and the authors variously choose to use empirical or theoretical derivations for the zero-point.  In all cases
the error on the resulting distance modulus is $\sim 0.1 - 0.15$ mag, and the range 18.48  - 18.58 mag.  However, with the new parallaxes 
from \citet{ben11} it is clear that using their absolute magnitude scale is the most robust way to establish the Population II distance scale at present; we
quote their results for the LMC distance taking the mean of their two methods for bias corrections, Lutz-Kelker-Hanson (LKK) \citep{han79} and
reduced parallaxes (RP), \citep{fea02}, however the difference in LMC modulus between these is small, and they find that an RR Lyrae star with
$[Fe/H] = -1.5$ has $M_{V} = 0.45 \pm 0,05$ (LKK) or $M_{V} = 0.46 \pm 0.03$ (RP).  This is substantially brighter than the derived value of
$M_{V} = 0.66 \pm 0.14$ at $[Fe/H] = -1.48 \pm 0.07$ \citep{cat08} based on RR Lyrae alone, but with much smaller error. The results given
below would indicate the analyses of $K$ band data based on RR Lyrae need to have their derived LMC modulus adjusted by only a small 
amount, $\sim 0.02$ mag when the new calibration is applied.   The reasons for the smaller adjustment in the IR will be of interest to investigate.

The distance to the LMC from analyses of field RR Lyraes observed in optical band passes depends on the treatment of the reddening; this is 
not only a problem just for the RR Lyraes of course and is probably the strongest driver for moving to the infrared when at all possible. Also,
three of the five galactic RR Lyrae stars with HST parallaxes have $E(B-V) \geq 0.1$ mag, so assuming the $K$ magnitudes for the parallax stars are accurate (see
\citet{fea08} for a description of how the $K$ band magnitudes were obtained), then the zero-point is potentially more robust in the infrared as well.
Thus in the optical, the primary difference between distances derived from the data of \citet{gra04} and  \citet{sos03} is due to the different
reddening used in the two studies, with \citet{ben11} finding an LMC modulus of $18.61 \pm 0.05$ for the \citet{gra04} data and $18.46 \pm 0.06$
for the \citet{sos03} data.   In the infrared, they quote $18.55 \pm 0.05$ for the K band data analyzed by \citet{bor09} and $18.50 \pm 0.03$  for
the RR Lyraes in the Reticulum cluster measured by \citet{dal04}.   

For the SMC, a recent Fourier analysis of the light curves for 335 RRab and 17 RRc selected from an OGLE-II sample of 536 single-mode RR Lyraes
by \citet{deb10} yields a distance modulus of 
$18.83 \pm 0.01 $ mag (internal) using intensity-weighted mean magnitudes and an absolute magnitude calibration from \citet{kov96}.  It would be
interesting to compare this result with that obtained with a more modern RR Lyrae zero-point calibration (some 0,2 mag brighter) and re-examination 
of the reddening corrections,  \citet{kap11} similarly analyze 100 carefully selected RR Lyraes from the OGLE-II and OGLE-III catalogs, and calculate 
an SMC modulus of $18.90 \pm 0.18$ referenced to an LMC modulus of $18.52 \pm 0,06$. The RR Lyraes from the OGLE catalogs are clearly a very 
rich resource for delineating the structure of the MC as exhibited by its oldest populations, in addition to mean distances to the galaxies
themselves.     

We conclude that the RR Lyrae results for the LMC distance tend in the mean to give a result near $18.53$ mag, with more dispersion than is desirable
in the optical analyses, due to differing treatments of reddening.  Further analysis of the OGLE data together with the new Galactic calibration, 
should allow the situation to be dramatically improved.

\section{Red Clump Stars}

Red clump stars are common in the solar vicinity and many have accurate Hipparcos parallaxes allowing an accurate mean absolute magnitude for the solar
sample to be derived \citep{alv00,gro08}.  Several authors e.g. \citet{per03} and references therein have shown that substantial corrections
are needed for age and metallicity, and that these are very dependent on the band pass, for instance if the stars are around 2 Gyr old then the
$K$ band is excellent with no metallicity correction needed, whereas if the stars are very old the $I$ band is better, and generally the $V$ band 
should be avoided \citep{sal03}. This conclusion was also reached empirically by \citet{pie03}.  The theoretically derived population 
corrections and in particular the K band mean absolute magnitude found by \citet{sal03} are in excellent agreement with the revised Hipparcos parallaxes 
for red clump stars \citep{gro08}. Thus, \citet{sal03} find a distance modulus for the LMC of $18.47 \pm 0.01$ (random) $\pm 0.06$ (systematic) using the
LMC photometry by \citet{pie03} who from their own analysis and the \citet{alv00} calibration find an LMC modulus of $18.50 \pm 0.01$ (internal), while
analysis of an independent data set by \citet{koe09} finds $18.54 \pm 0.06$. However, measuring red clump $K$ magnitudes for 17 intermediate age LMC
clusters, \citet{gro07} find a rather shorter mean distance,  $18.40 \pm 0.04$ (random) $\pm 0.08$ (systematic).  
 
In summary, there is now deep understanding on how to determine distances from red clump stars, and providing the caveats are respected the method can be very reliable
as the stars are relatively common.  The results cluster around an LMC modulus of $18.48$.

\section{Eclipsing Binaries}

Detached binaries that are spectroscopically double-lined and also photometrically eclipsing allow the measurement of the fundamental 
stellar parameters for the two companions, with the distance able to be determined with the addition of the absolute surface brightness, first
pointed out by \citet{lac77}. The absolute surface brightness 
can be found empirically from the surface brightness - color relation which is well established for stars of spectral type later than A5 from 
interferometrically-measured stellar angular diameters.  For early type stars the process is less empirical and at present relies on accurately 
determining the effective temperatures e.g. \citet{bon11}.  But in principle, the use of detached double-lined eclipsing binaries for determining
distances with a minimum of assumptions \citep{and91,tor10} has the potential to allow one-step distances to several nearby Local Group galaxies
\citep{pac96,hil06}.  Until very recently this type of analysis 
in the LMC has been restricted
to a few early type systems \citep{gui99,rib02,fit02,fit03}, and to 21 detached systems in the SMC \citep{hil05}.  The first LMC results tended to
give rather short distances, $18.30 \pm 0.07$ \citep{gui99}, $18.38 \pm 0.08$ 
\citep{rib02},  $18.53 \pm 0.05$ \citep{fit02},  $18.18 \pm 0.08$ \citep{fit03}.  The latter system has low (galactic only) reddening which would 
support it being foreground to the main body of the LMC, as suggested by \citet{fit03}.
The faintness of late type systems at the distance of the MC makes their study difficult, however    
OGLE III discoveries of G giant systems will enhance the sample of LMC detached eclipsing binaries
considerably and allow 
a more direct derivation of distance than with the early type systems, with smaller expected systematic uncertainty.
The first of these new analyses, for a system consisting of two G4III stars, has been published by \citet{pie09}. Their
derived value for the distance of this system is $18.50 \pm 0.06$ mag, using the $V-K$ surface brightness - color relation of \citet{dib05}.  \citet{pie10}
point out that with the discovery of a few dozen such systems in the MC we will be able to determine the distances to both galaxies to $1-2\%$ accuracy
from this method alone.

In summary, unlike the other LMC distance indicators discussed above that due to their copious numbers can have the statistical errors in their
mean apparent distance reduced to small levels, there are at present only  six detached double-lined eclipsing binaries in the LMC with 
published distances.  Three of these
cluster closely around a modulus of $18.52$ mag, the other three have much shorter moduli.  Setting aside the very discrepant value for HV5936 ($18.18$ mag),
 the remaining five systems provide a mean LMC modulus of $18.45$ mag.

\section{Conclusions}

The distance indicators discussed above show remarkable agreement.  The mean of the mean distances stated above at the end of each section 
for the five distance indicators is $18.48$ mag, and the range is $\pm 0.05$ mag. With most of the measurements discussed quoting internal
errors typically $0.01-0.03$ mag, and with the inclusion of external (calibration) errors typically $0.04 - 0.09$ mag, adopting the mean of
the means and the range as our present best estimate for the distance of the LMC and its error seems justifiable.  The Cepheid result (18.45 mag)
is uncorrected for the metallicity difference between the Galaxy and the LMC; a corollary is that such a correction cannot be very large.

An accurate distance scale is a prerequisite for studying many of the most profound question in astronomy,and the MC.
particularly the LMC, continue to play a pivotal role in improving and understanding a variety of distance indicators.  Major recent surveys
of the MC have provided enormous increases in the numbers of such stars, helping to improve our understanding of their
properties and allowing the preparation of unbiased samples for comparison in detail with stars in our own Galaxy and elsewhere.  The sample
sizes are such that calibration and cross-calibration effects are becoming dominant, however the demands of the target science are such that
these effects need to be understood at the $\sim 0.02 $ mag level, and preferably at least a factor of two better. The several standard
systems in the near IR require very careful transformation work, and similar procedures will be needed in the optical, where the tendency for 
the new wide field instruments becoming available in the south on large telescopes to survey in the filters approximating the SDSS system
will complicate comparisons
with legacy data obtained in the Johnson-Cousins system. 

The zero-point calibrations have improved for both Cepheids and RR Lyraes.  Hipparcos first provided parallaxes for Cepheids and
red clump stars, while HST FGS has provided high quality measurements for a few Cepheids and RR Lyraes, and should soon also 
further strengthen the Population II scale with good parallaxes for a few subdwarfs.  GAIA will make
dramatic improvements with factors of at least a hundred increase in both sample sizes and astrometric accuracy.  
Eclipsing binaries also provide a very fundamental way to determine the distances to the MC, and here 
the discovery of G III examples, allowing a more straightforward analysis, is a major advance.  Improved calibrations and 
analysis techniques for the early type stars are also providing more reliable results for these stars.   

Finally, intense efforts on understanding the evolutionary properties of post main sequence turnoff stars, and in particular 
the properties of pulsating variables, has placed 
analysis of the observational results on a very sound theoretical footing.  Although there are still important effects that elude us, such as the luminosity
dependence of Cepheids on metallicity, and determining how pervasive is the need to vary helium content, there has been substantial progress, with
the red clump stars now a well-understood distance indicator being a prime example of this. However, we are not home yet.  The desire to know the Hubble 
Constant to $\sim 1\%$ places great demands on both observations and theory, it will require extraordinary care to improve
the accuracy of the present result by another factor of two or three.


\begin{thebibliography}{}
\bibitem[Alves (2000)]{alv00} Alves, D.R. 2000, \apj, 539, 732 
\bibitem[Alves(2005)]{alv05} Alves, D.R. 2005, Highlights in Astronomy, IAU, O. Engvold, San Francisco, CA: Astronomical Society of the Pacific, Vol. 13, 1448
\bibitem[Andersen (1991)]{and91} Andersen, J. 1991, \aapr, 3, 91
\bibitem[Benedict et al. (2002)]{ben02} Benedict, G.F., McArthur, B.E., Fredrick, L.W., et al. 2002, \aj, 124, 1695
\bibitem[Benedict et al. (2007)]{ben07} Benedict, G.F., McArthur, B.E., Feast, M.W., et al. 2007, \aj, 133, 1810
\bibitem[Benedict et al.(2011)]{ben11} Benedict, G.~F., McArthur, B.~E., Feast, M.W., et al. 2011, \aj, 142, 187
\bibitem[Besla et al. (2011)]{bes11} Besla, G., Kallivayalil, N., Hermquiat, L., et al. 2010, \apj, 721, L97
\bibitem[Bonanos et al. (2011)]{bon11} Bonanos, A.Z., Castro, N., Macri, L.M., Kudritzki, R-P. 2011, \apj, 729, L9
\bibitem[Bono et al. (2003)]{bon03} Bono, G., Caputo, F., Castellani, V., et al. 2003, \mnras, 344, 1097 
\bibitem[Bono et al. (2008)]{bon08}  Bono, G., Caputo, F., Fiorentino, G, Marconi, M., \& Musella, I. 2008, \apj, 684, 102
\bibitem[Borissova et al. (2004)]{bor04} Borissova, J., Minniti, D., Rejkuba, M., et al. 2004, \aap, 423, 97
\bibitem[Borissova et al.(2009)]{bor09} Borissova, J., Rejkuba, M., Minniti, D., Catelan, M., \& Ivanov, V.~D.\ 2009, \aap, 502, 505 
\bibitem[Caldwell \& Coulson (1986)]{cal86} Caldwell, J.A.R., \& Coulson, I.M. 1986, \mnras, 218, 223
\bibitem[Caputo (2011)]{cap11} Caputo, F., 2011, this meeting 
\bibitem[Cardelli et al. (1989)]{car89} Cardelli, J.A., Clayton, G.G., \& Mathis, J.S. 1989, \aj, 96, 695
\bibitem[Carreta et al. (2000)]{car00} Carretta, E., Gratton, R.G., Clementini, G., \& Fusi Pecci, F. 2000, \apj, 533, 215
\bibitem[Catelan et al. (2004)]{cat04} Catelan, M., Pritzl, B.J., \& Smith, H.A. 2004, \apjs, 154, 633
\bibitem[Catelan \& Cortes (2008)]{cat08} Catelan, M., \& Cortes, C. 2008, \apj, 676, L135
\bibitem[Chaboyer et al.(2011)]{cha11} Chaboyer, B.~C., Benedict, G.~F., McArthur, B.~E., et al.\ 2011, Bulletin of the American Astronomical Society, 43, \#242.25 
\bibitem[Clementini et al. (2003)]{cle03} Clementini, G., Gratton, R.G., Bragaglia, A.,et al. 2003, \aj, 125, 1309
\bibitem[Clementini(2008)]{cle08} Clementini, G. 2008, IAU Symposium 256, J. Th. van Loon \& J.M. Oliveira, Cambridge: CUP, 373
\bibitem[Coppola et al. (2011)]{cop11} Coppola G., Dall'Ora, M., Ripepi, V., et al. 2011, \mnras, 416, 1056
\bibitem[Dall'Ora et al. (2004)]{dal04} Dall'Ora, M., Storm, J., Bono, G., et al. 2004, \apj, 610, 269
\bibitem[Deb \& Singh (2010)]{deb10} Deb, S., \& Singh, H.P. 2010, \mnras, 402, 691
\bibitem[di Benedetto (2005)]{dib05} di Benedetto, G.P. 2005, \mnras, 357, 174
\bibitem[Di Criscienzo et al. (2004)]{dic04} Di Criscienzo, M., Marconi, M., \& Caputo, F. 2004, \apj, 612, 1092 
\bibitem[Feast (2002)]{fea02} Feast, M.W. 2002, \mnras, 337, 1035
\bibitem[Feast et al. (2008)]{fea08} Feast, M.W., Laney, C.D., Kinman, T.D., Van Leeuwen, F., \& Whitelock, P.A. 2008, \mnras, 386, 2115
\bibitem[Feast \& Catchpole (1997)]{fea97} Feast, M.W. \& Catchpole, R.F. 1997, \mnras, 286, L1
\bibitem[Feast \& Walker (1987)]{few87} Feast, M.W. \& Walker, A.R. 1987, \araa, 25, 345
\bibitem[Fitzpatrick et al. (2002)]{fit02} Fitzpatrick, E.F., Ribas, I., Guinan, E.F., et al. 2002, \apj, 564, 26-
\bibitem[Fitzpatrick et al. (2003)]{fit03} Fitzpatrick, E.F., Ribas, I., Guinan, E.F., Maloney, F.P., Claret, A. 2003, \apj, 587, 685
\bibitem[Fouqu\'{e} et al. (2007)]{fou07} Fouqu\'{e}, P., Arriagada, P., Storm, J., et al. 2007, \aap, 476, 73
\bibitem[Freedman \& Madore (2009)]{fre09} Freedman, W.L., \& Madore, B.F. 2009, \apj, 696, 1498
\bibitem[Freedman \& Madore (2010)]{fre10} Freedman, W.L. \& Madore, B.F. 2010, \araa, 48, 673
\bibitem[Freedman \& Madore (2011)]{fre11} Freedman, W.L. \& Madore, B.F. 2011, \apj, 734, 46
\bibitem[Gieren et al. (2005)]{gie05} Gieren, W., Storm, J., Barnes T.G., et al. 2005, \apj, 627, 224
\bibitem[Gratton et al. (2004)]{gra04} Gratton, R.G., Bragaglia, A., Clementini, G., et al. 2004, \aap, 421, 937
\bibitem[Grocholski et al. (2007)]{gro07} Grocholski, A., Sarajedini, A., Olsen, K.A.G., Tiede, G.P., \& Mancone, C.L. 2007, \aj, 134, 680
\bibitem[Groenwegen (2008)]{gro08} Groenwegen, M. 2008, \aap, 488, 935
\bibitem[Guinan et al. (1998)]{gui99} Guinan, E.F., Fitzpatrick, E.L., Dewarf, L.E., et al.1998, \apj, 509, L21 
\bibitem[Hanson (1979)]{han79} Hanson, R.P. 1979, \mnras, 186, 875
\bibitem[Hilditch et al. (2005)]{hil05} Hilditch, R.W., Howarth, I.D., \& Harries, T.J. 2005, \mnras, 357, 304
\bibitem[Hilditch (2006)]{hil06} Hilditch, R.W. 2006, \apss, 304, 203
\bibitem[Hislop et al. (2011)]{his11} Hislop. L., et al. 2011, \apj, 733, 79
\bibitem[Ita \& Matsunaga (2011)]{ita11} Ita, Y. \& Matsunaga, N. 2011, \mnras, 412, 2345
\bibitem[Kapakos et al. (2011)]{kap11} Kapakos, E., Hatzidimitriou, D., \& Soszy\'{n}ski, I. 2011, \mnras, 415, 1366
\bibitem[Koerwer (2009)]{koe09} Koewer, J.F. 2009, \aj, 138,1 
\bibitem[Komatsu et al. (2011)]{kom11} Komatsu,E., Smith, K.M., Dunkley, J., et al. 2011, \apjs, 192, 18
\bibitem[Kov\'{a}cs \& Jursik (1996)]{kov96} Kov\'{a}cs, G. \& Jursik, J. 1996, \apj, 466, 47
\bibitem[Kudritzki et al. (2008)]{kud08} Kudritzki, R-P., Urbaneja, M.A., Bresolin, F., et al. 2008, \apj, 681, 269
\bibitem[Lacey (1977)]{lac77} Lacey, C. 1977, \apj, 213, 458
\bibitem[Macri et al.(2006)]{mac06} Macri, L.M., Stanek, K.Z., Bersier, D., Greenhill, L.J., \& Reid, M.J. 2006, \apj, 652, 1133
\bibitem[Madore \& Freedman (2009)]{mad09} Madore, B.F. \& Freedman, W.L. 2009, \apj, 696, 1498
\bibitem[Majaess et al. (2011)]{maj11} Majaess, D., Turner, D. \& Gieren, W. 2011, \apj, 741. L36
\bibitem[Mould \& Sakai (2008)]{mou08} Mould, J., \& Sakai, S. 2008, \apj, 686, 75
\bibitem[Mould \& Sakai (2009a)]{mou109} Mould, J., \& Sakai, S. 2009a, \apj, 694, 1331
\bibitem[Mould \& Sakai (2009b)]{mou209} Mould, J., \& Sakai, S. 2009b, \apj, 697, 996
\bibitem[Nemec et al. (2009)]{nem09} Nemec, J.N., Walker, A.R., \& Jeon, Y-B. 2009, \aj, 138, 1310
\bibitem[Nidever et al. (2010)]{nid10} Nidever, D.L., Majewski, S.R., Butler Burton, W., \& Nigra, L. 2010, \apj, 723, 1618
\bibitem[Nikolaev et al. (2004)]{nik04} Nikolaev, S., Drake, A.J., Keller, S,C., et al. 2004, \apj, 601,260
\bibitem[Olsen et al. (2011)]{ols11} Olsen, K.A.G., Zaritsky, D, Blum, R.D., Boyer, M.L., \& Gordon, K.D. 2011, \apj, 737, 29
\bibitem[Paczynski (1996)]{pac96} Paczynski, B. 1996, Extragalactic Distance Scale STScI Symp., arXiv:astro-ph/9608094
\bibitem[Percival \& Salaris (2003)]{per03} Percival, S., \& Salaris, M. 2003, \mnras, 343, 539
\bibitem[Pejcha \& Stanek (2009)]{pej09} Pejcha, O., \& Stanek, K.Z. 2009, \apj, 704, 1730
\bibitem[Pietrzy\'{n}ski et al. (2003)]{pie03} Pietry\'{n}ski, G., Gieren, W, \& Udalski, A 2003, \aj, 125, 2494
\bibitem[Pietrzy\'{n}ski et al (2009)]{pie09} Pietry\'{n}ski, G., Thompson, I.B., Graczyk, D., et al. 2009, \apj, 697, 862
\bibitem[Pietrzy\'{n}ski et al. (2010)]{pie10} Pietry\'{n}ski, G., Thompson, I., Graczyk, D., Gieren, W., \& Minniti, D. 2010, ASP Conf. Ser., 435, 13
\bibitem[Ribas et al. (2002)]{rib02} Ribas, I., Fitzpatrick, E.L., Maloney, F.P., Guinan, E.F., Udalski, A. 2002, \apj, 574, 771
\bibitem[Riess et al. (2011)]{rie11} Riess, A.G., et al., 2011, \apj, 730, 119
\bibitem[Romaniello et al. (2009)]{rom09} Romaniello, M., Primas, F., Mottini, M., et al. 2009, AIPC, 1170, 99
\bibitem[Saha et al. (2010)]{sah10} Saha, A., Olszewski, E.W., Brondel, B., et al. 2010, \aj, 140, 1719
\bibitem[Sakai et al. (2004)]{sak04} Sakai, S., Ferrarese, L., Kennicut, R.C., \& Saha, A. 2004, \apj, 608, 42
\bibitem[Salaris et al. (2003)]{sal03} Salaris, M., Percival, S., \& Girardi, L. 2003, 345, 1030
\bibitem[Schaefer (2008)]{sha08} Schaefer, B.E. 2008, \aj, 1354,112
\bibitem[Scowcroft et al.(2009)]{sco09} Scowcroft, V., Bersier, D., Mould, J.R., \& Wood, P.R. 2009, \mnras, 396, 1287
\bibitem[Sollima et al. (2006)]{sol06} Sollima, A., Cacciari, C., \& Valenti, E. 2006, \mnras, 372, 1675 
\bibitem[Soszy\'{n}ski et al. (2002)]{sos02} Soszy\'{n}ski, I., Udalski, A., Szyma\'{n}ski, M.,  et al. 2002, \actaa, 52, 369
\bibitem[Soszy\'{n}ski et al. (2003)]{sos03} Soszy\'{n}ski, I., Udalski, A., Szyma\'{n}ski, M., et al. 2003, \actaa, 59, 1
\bibitem[Soszy\'{n}ski et al.(2004)]{sos04} Soszy\'{n}ski, I., Udalski, A., Kubiak, M., et al.\ 2004, \actaa, 54, 129 
\bibitem[Soszy\'{n}ski et al. (2009)]{sos09} Soszy\'{n}ski, I., Udalski, A., Szyma\'{n}ski, M., et al. 2009, \aap, 59, 239
\bibitem[Storm et al. (2011a)]{sto11} Storm, J., Gieren, W., Fouqu\'{e},P., 2011a, \aap, 534, 94
\bibitem[Storm et al. (2011b]{sto211} Storm, J., Gieren, W., Fouqu\'{e},P., 2011b, \aap, 534, 95
\bibitem[Szewczyk et al. (2008)]{sze08} Szewczyk, O., Pietrzy\'{n}ski, G., Gieren, W., et al. 2008, \aj, 136, 272
\bibitem[Tabur et al.(2010)]{tab10} Tabur, V., Bedding, T.~R., Kiss, L.~L., et al.\ 2010, \mnras, 409, 777 
\bibitem[Tammann et al. (2008)]{tam08} Tammann, G.A., Sandage, A., \& Reindl, B. 2008, \apj, 679, 52
\bibitem[Torres et al. (2010)]{tor10} Torres, G, Andersen, J., \& Gim\'{e}nez, A. 2010, \aapr, 18, 67
\bibitem[Turner (2010)]{tur10} Turner, D.G. 2010, \apss, 326, 219
\bibitem[Turner \& Burke (2002)]{tur02} Turner, D.G. \& Burke, J.F. 2002, \aj, 124, 2931
\bibitem[van Leeuwen et al. (2007)]{van07} van Leeuwen, F., Feast, M.W., Whitelock, P.A., \& Laney, C.D. 2007, \mnras, 379, 723
\bibitem[Walker(2003)]{wal03} Walker, A.R. 2003, Stellar Candles for the Extragalactic Distance Scale,  Lecture Notes in Physics, D. Alloin \& W. Gieren, 635, 265
\bibitem[Whitelock \& Feast(2000)]{whi00} Whitelock, P.A. \& Feast, M.W. 2000, \mnras, 319, 759
\bibitem[Whitelock et al.(2008)]{whi08} Whitelock, P.A., Feast, M.W. \& van Leeuwen, F. 2008, \mnras, 386, 313
\bibitem[Windmark et al. (2011)]{win11} Windmark, F., Lindgren, L., Hobbs, D. 2011, \aap, 530, 76
\bibitem[Wood(2000)]{woo00} Wood, P.~R.\ 2000, \pasa, 17, 18 
\end{thebibliography}
\end{document}